# Hydrogen partial pressure as a key parameter in the reactive sputtering of photochromic YHO thin films

E. Strods[1*], M. Zubkins[1], E. Letko[1], V. Vibornijs[1], A. Sarakovskis[1], J. Purans[1]

[1]Institute of Solid State Physics, University of Latvia, Kengaraga 8, LV-1063, Riga, Latvia

[*]Corresponding author: edvards.strods@cfi.lu.lv

**Abstract**

Oxygen-containing yttrium hydride (YHO) thin films were synthesised by reactive magnetron sputtering over a wide range of hydrogen-to-argon gas flow ratios ($H_2$/Ar) during yttrium hydride deposition, followed by post-oxidation in air. The deposition process was monitored in situ by plasma optical emission spectroscopy. X-ray diffraction reveals a strong dependence of crystallographic texture on the $H_2$/Ar ratio: highly <100>-textured films formed at low $H_2$/Ar ratios progressively evolve into more polycrystalline microstructures with slightly reduced crystallite size as the hydrogen content increases. The optical transmittance gradually decreases from approximately 85 % to 70 % as the $H_2$/Ar ratio is raised to 0.6, followed by a sharp drop at higher hydrogen contents. The photochromic contrast of 705–845 nm thick YHO films, measured after 10 min of illumination with a solar simulator, decreases from 14 % to 7 % as the $H_2$/Ar ratio increases from 0.08 to 0.45. Thinner films (365–505 nm) exhibit a similar but less pronounced trend. Bleaching kinetics, analysed using a double-exponential model, indicate faster recovery, with effective bleaching time constants of approximately 1–15 min for films deposited at $H_2/Ar < 0.25$. X-ray photoelectron spectroscopy reveals higher oxygen incorporation in films deposited at lower $H_2$/Ar ratios, deviating from the commonly reported correlation between oxygen content and photochromic response in YHO. These results demonstrate that low hydrogen partial pressure during reactive sputtering is advantageous for achieving higher optical transparency, enhanced photochromic contrast, and faster bleaching kinetics in YHO thin films.

## 1. Introduction

Oxygen-containing rare-earth (RE) hydride thin films (REHOs) have emerged as a promising class of inorganic photochromic materials for applications such as smart windows and adaptive optical coatings [1]. Among these materials, oxygen-containing yttrium hydride (YHO) [2], also referred to as yttrium oxyhydride, is of particular interest [3]. It exhibits (i) optical transmittance of 70–80 % in the clear state, (ii) colour-neutral photodarkening with contrast values reaching 40–50 % typically achieved after prolonged illumination or when using high-intensity or high-energy light sources, and (iii) reversible bleaching under ambient dark conditions [4]. These properties can be achieved using scalable roll-to-roll deposition techniques [5].

YHO thin films are typically prepared by depositing $\beta$-$YH_x$ ($x$ = 1.8–2.3 [6,7]) via reactive (Ar + $H_2$) magnetron sputtering, followed by oxidation upon exposure to an oxygen-containing atmosphere [8]. During oxidation, oxygen partially replaces hydrogen in tetrahedral lattice sites [9,10], leading to the formation of a transparent, photochromic oxyhydride phase [3]. This process involves both hydrogen release and rearrangement, resulting in changes to the electronic structure [11], as confirmed by extended X-ray absorption fine structure (EXAFS) studies [12]. In YHO, photochromic behaviour is observed over a relatively broad compositional range, however, pronounced photochromic contrast is typically achieved only within a narrower window, corresponding to an O/Y atomic ratio of approximately 0.5–1.0 [4].

Despite extensive research, the mechanisms underlying photochromism in YHO remain an active topic of debate. Proposed explanations include two principal mechanisms: the formation of metallic domains or anion vacancy related electronic transitions. Metallic domains may develop either through the reduction of trivalent yttrium by photo-excited electrons [13] or via anion diffusion processes [14–16], while the vacancy-based mechanism involves local changes in charge state and lattice structure [17]. The relatively slow bleaching kinetics observed in many YHO films indicate that the photochromic response cannot be explained solely by fast electronic processes, suggesting the involvement of mobile hydrogen [18–20]. Previous studies have demonstrated that both the magnitude and kinetics of the photochromic response depend strongly on factors such as chemical composition [21], anionic ordering within the sublattice [10], defect populations [13,17], film thickness [22], and illumination conditions [23]. Considering these dependencies, various strategies have been explored to enhance the performance of YHO-based films. These approaches include the optimization of deposition parameters, fabrication of multilayer coatings [24–27], doping [28,29], cation mixing [30], among others.

Reactive sputtering is a widely used technique for the deposition of compound thin films [31]. In this process, a metallic target is sputtered in an atmosphere containing argon and a reactive gas, which reacts with the sputtered species to form the desired compound on the substrate. The film composition and chemistry can be controlled by adjusting the reactive gas flow [32], while the use of metallic targets typically enables higher deposition rates and the operation of scalable direct current (DC) power supplies.

However, the reactive gas also poisons the target surface, altering its sputtering behaviour, typically reducing the deposition rate, and giving rise to a characteristic hysteresis loop that separates metallic, unstable transition, and compound modes [33]. As a result, reactive sputtering often requires precise process control, typically based on feedback from parameters such as discharge voltage, reactive gas partial pressure, or optical emission from the plasma, which are used to stabilise the process through adjustment of the reactive gas flow. In practice, a favourable balance between deposition rate and film composition must be achieved.

Reactive processes employing oxygen or nitrogen to deposit metal oxide and nitride films are well established, however, the reactive deposition of hydrides has been far less studied. In particular, the role of the hydrogen partial pressure during sputtering in determining the properties of the YHO films has not yet been clearly understood. In this study, we investigate the influence of the $H_2$/Ar flow ratio used during reactive magnetron sputtering on the structural and optical, including photochromic, properties of YHO thin films. Gaining such insight is essential for the transition of YHO sputtering from a laboratory scale process to a viable method for large-area industrial scale production.

## 2. Experimental details

YHO thin films were synthesized on soda-lime glass substrates by reactive pulsed-DC magnetron sputtering (80 kHz, 2.5 μs off-time) by a PVD coater (G500M, Sidrabe Vacuum, Ltd). A 99.9% pure yttrium target (150 mm × 75 mm × 3 mm) was sputtered in a mixed Ar (99.9999 %) and $H_2$ (99.999 %) atmosphere using a planar balanced magnetron. The hydrogen-to-argon gas flow ratio ($H_2$/Ar) was varied between 0.01 and 1.00 while keeping the argon flow constant, resulting in a working pressure increase from 0.71 to 0.89 Pa. Depositions were carried out at a constant time-averaged power of 200 W, with no intentional substrate heating. Film thickness was controlled by adjusting the deposition time (10–45 min). Following deposition, the $\beta$-$YH_x$ films were deliberately exposed to air to form the YHO phase.

Plasma optical emission spectra (OES) from the discharge were collected using an optical fibre probe positioned inside the chamber, parallel to the target surface and approximately 1 cm above it. Time-averaged OES were recorded at 0.3 s intervals using a PLASUS EMICON MC spectrometer, with the probe position and spectrometer settings kept unchanged for all deposition runs. Before each deposition, the Y target was pre-sputtered at 200 W and 0.40 Pa, in pure Ar until the Y emission intensity and target voltage reached stable values. The working pressure was then adjusted to 0.71 Pa, and, prior to $H_2$ introduction, the Y ($I_{\mathrm{Y,m}}$) and Ar ($I_{\mathrm{Ar,m}}$) emission intensities were recorded and used as run-specific normalization references. Hydrogen was subsequently introduced, and after the discharge had stabilised, deposition was started and the emission intensities of Y ($I_{\mathrm{Y}}$), Ar ($I_{\mathrm{Ar}}$), and H ($I_{\mathrm{H}}$) were recorded. The H emission intensity was normalised to the Ar emission intensity.

The crystallographic structure of the films was examined by X-ray diffraction (XRD), using a Rigaku MiniFlex600 X-ray powder diffractometer with Bragg-Brentano $\theta$-2$\theta$ geometry and a 600 W Cu anode (Cu Kα radiation, $\lambda$ = 1.5406 Å) X-ray tube. Crystallite size was estimated using the Scherrer equation, $\tau = K\lambda\,(\beta\cos\theta)^{-1}$. Microstrain contributions were not subtracted, consequently, the reported crystallite sizes should be regarded as first-order estimates [34]. The degree of preferential orientation was evaluated using the texture coefficient ($TC$), calculated according to equation:

$$TC_{hkl} = \frac{I_{hkl}/I^*_{hkl}}{\frac{1}{n}\sum(I_{hkl}/I^*_{hkl})}, \tag{1}$$

where $I_{hkl}$ is the experimental intensity of the XRD peak corresponding to the (*hkl*) plane, $I^*_{hkl}$ is the reference intensity for the same plane obtained from the ICDD database for $\beta$-$YH_2$ (ICDD 04-002-6939), and $n$ is the number of reflections used in the calculation. For all samples, the texture coefficient was calculated using the 111, 200, and 311 reflections ($n$=3). The 400 reflection was excluded because it was not consistently observed across the entire sample series.

XPS measurements were performed by an ESCALAB 250Xi (ThermoFisher) instrument to determine the chemical composition of YHO films. An Al Kα X-ray tube with energy 1486 eV was used as an excitation source, the size of the analysed sample area was 650 × 100 µm, the angle between the analyser and the sample surface was 90°. An electron gun was used to perform charge compensation. The base pressure, with the charge neutralizer switched on during spectra acquisition, was better than $10^{-7}$ mbar. The films were gently sputter-cleaned by $Ar^+$ cluster gun with the ion energy 6.0 keV (150 Ar atoms) for 10 s before the measurements. The size of the cleaned area was 2 × 2 mm. The pass energy of the measured spectra was 20 eV. The calibration and linearity of the binding energy scale was confirmed by measuring the positions of Ag $3d_{5/2}$, Au $4f_{7/2}$ and Cu $2p_{3/2}$ to be at 368.21 eV, 83.93 eV and 932.58 eV, respectively. The full width at half maximum (FWHM) of Au $4f_{7/2}$ peak was better than 0.58 eV. The binding energy spectra of all the samples were referenced against the adventitious carbon peak to appear at 284.8 eV.

Film thicknesses were determined using a SE WOOLLAM RC2 spectroscopic ellipsometer in the spectral range of 2.25–0.73 eV, where the films exhibit high transparency. The Cauchy model, $n(\lambda) = A + B\lambda^{-2}$, was applied in this range for the analysis. The primary ellipsometric angles, Ψ and Δ, were measured at incident angles ranging from 55° to 65° in 5° increments. Spectroscopic ellipsometry (SE) experimental data and model-based regression analyses were processed using the WOOLLAM software, CompleteEASE. The mean squared error (MSE) values for the models ranged between 10 and 38, with most samples falling between 10 and 15. The thickness of the YHO films in this study ranged from 255 to 860 nm. A detailed analysis of the SE measurements over the broader spectral range of 5.9–0.7 eV, employing physical oscillator models, along with the optical constants and gradients is provided in Ref. [3].

The photochromic response was tested under ambient conditions using an Oscilla LED solar simulator (100 mW/cm$^2$) as the illumination source. A tungsten-halogen lamp with the power density of ≈3.3 mW/cm$^2$ was used for optical probing. No measurable photo-darkening was induced by the probing light. A schematic of the measurement setup, along with the emission spectra of the light sources used, see Fig. S1 in SI (Supplementary Information - SI).

The optical transmittance and reflectance of the films in the range 250–2500 nm were measured by an Agilent Cary 7000 spectrophotometer. The sample was placed at an angle of 6° against the incident beam of *P*-polarized light, and the detector was placed at an angle of 180° behind the sample to measure the transmittance and at 12° in front of the sample to measure the specular reflectance. The luminous ($T_{lum}$) and solar ($T_{sol}$) were obtained from the transmittance spectra using the following expression:

$$T_{\mathrm{sol}} = \frac{\int T(\lambda)\varphi_{\mathrm{sol}}(\lambda)d\lambda}{\int \varphi_{\mathrm{sol}}(\lambda)d\lambda}, \tag{2}$$

where $\varphi_{sol}(\lambda)$ describes the solar irradiance spectrum at the Earth's surface. $T_{lum}$ was determined in the same way by replacing $\varphi_{sol}(\lambda)$ with $\varphi_{lum}(\lambda)$, representing the spectral sensitivity of the human eye [35].

## 3. Results and discussion

### *3.1. Reactive deposition of YHO thin films*

The time-averaged OES from the discharge during reactive sputtering is shown in Fig. 1a, and the emission lines of excited Ar (751 nm), Y (464 nm), and H (656 nm) were monitored during each deposition (Fig. 1b). The dependence of the normalised emission intensities $I_Y/I_{Y,m}$, $I_{Ar}/I_{Ar,m}$, and $I_H/I_{Ar}$, together with the discharge voltage and current, on the $H_2$/Ar flow ratio is presented in Fig. 1c. This normalisation accounts for variations in absolute emission intensity between deposition runs, allowing direct comparison across the investigated $H_2$/Ar ratios.

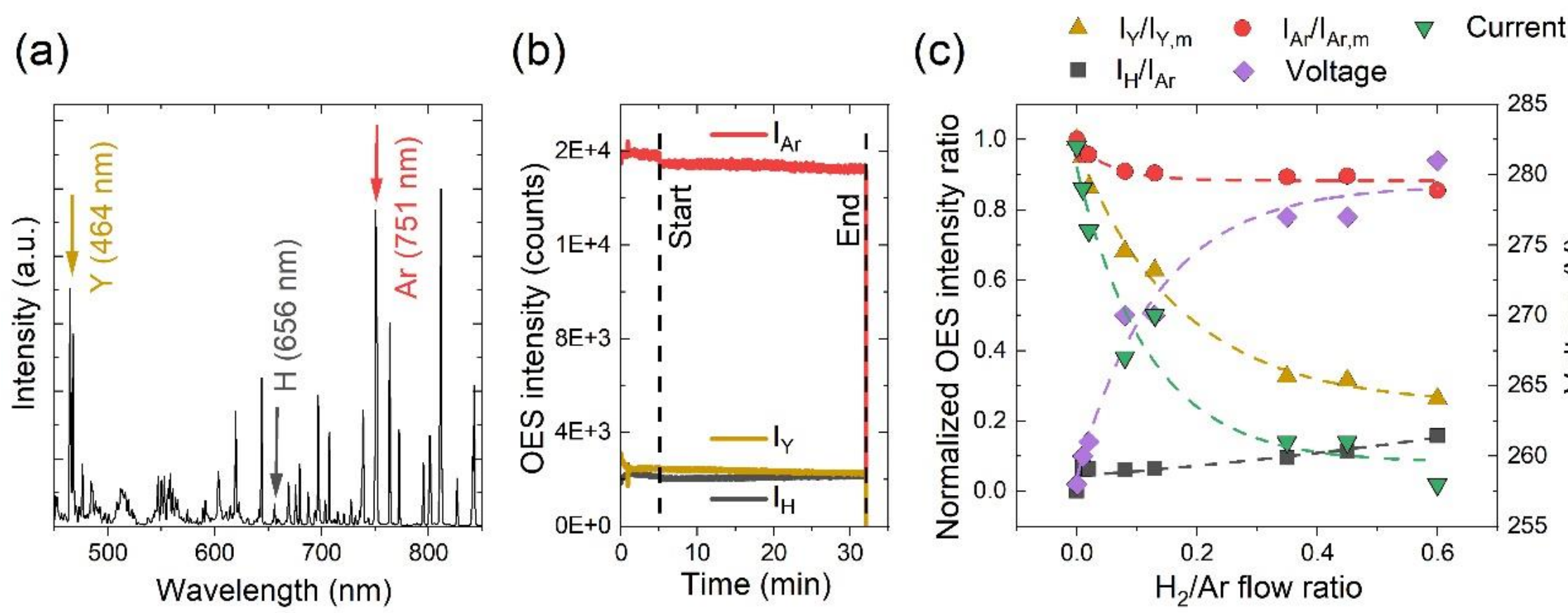

**Fig. 1.** (a) Representative optical emission spectrum measured during pulsed-DC magnetron sputtering of a Y target at an $H_2$/Ar flow ratio of 0.13, with characteristic emission lines of Y, H, and Ar indicated. (b) Time evolution of the emission intensities of Y, H, and Ar during deposition at $H_2/Ar = 0.6$; dashed vertical lines mark the start and end of the deposition period. (c) Normalised OES intensity ratios $I_Y/I_{Y,m}$, $I_{Ar}/I_{Ar,m}$, and $I_H/I_{Ar}$, together with the discharge voltage and current, plotted as functions of the $H_2$/Ar flow ratio.

Increasing the $H_2$/Ar ratio progressively hydridises the Y target, leading to a substantial reduction in sputtering yield. It is evidenced by the strong decrease in Y emission intensity with increasing $H_2$ flow (Fig. 1c), while excited argon emission decreases only slightly. Consequently, the deposition rate decreases by approximately a factor of 2.5 as the $H_2$/Ar ratio is increased from 0.01 to 0.60. Simultaneously, under constant power operation, the discharge voltage increases (Fig. 1c), reflecting an increase in plasma impedance due to reduced secondary electron emission from the target surface, either because of hydridisation and/or limited hydrogen ion bombardment, which produces fewer secondary electrons. A slight decrease in the argon ionisation efficiency leads to a modest decrease in the ion flux reaching both the target and the growing film. It is noted that hydrogen species ejected from the target during sputtering are expected to possess higher kinetic energies than thermalised hydrogen molecules in the gas phase.

The overall contribution of hydrogen ionisation to the discharge current is expected to be minor, as indicated by the low H emission intensity, and therefore cannot account for the observed substantial decrease in the deposition rate. Although the ionisation energy of atomic hydrogen is lower (13.6 eV) than that of argon (15.8 eV), molecular hydrogen must first undergo dissociation and its electron-impact ionisation cross-section ($6.6 \times 10^{-17}$ cm$^2$) is approximately four times smaller than that of argon ($2.8 \times 10^{-16}$ cm$^2$) [36]. Increasing the $H_2$ flow results in rise of total pressure from 0.71 to 0.89 Pa. The modest pressure increase associated with hydrogen addition is unlikely to significantly affect adatom mobility, energy transfer, or the angular distribution of incoming species through gas-phase collisions, owing to the low mass of hydrogen.

### *3.2. Structure of YHO thin films*

The XRD patterns of YHO films deposited as a function of thickness at a constant $H_2$/Ar flow ratio of 0.08, together with films grown at varying $H_2$/Ar ratios for two thickness ranges (365–505 nm and 705–845 nm), are shown in Fig. 2. All diffraction patterns reveal a face-centred cubic (*fcc*) Y cation sublattice characteristic of the YHO phase. The diffraction peaks are systematically shifted towards lower angles compared with $\beta$-$YH_2$, indicating an expanded lattice parameter due to oxygen incorporation, in agreement with previous reports [37–39].

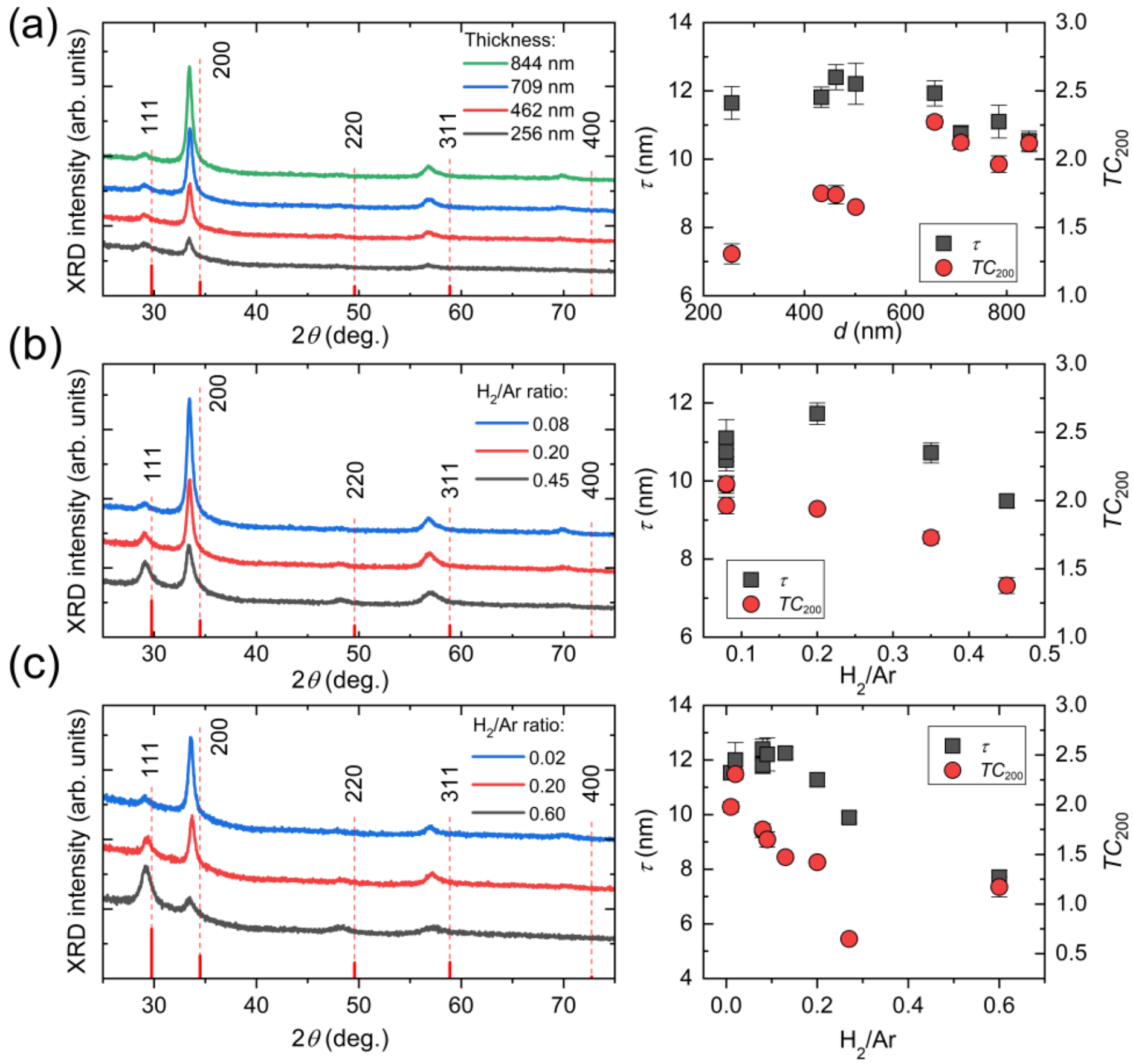

**Fig. 2.** X-ray diffraction (XRD) patterns of magnetron-sputtered YHO thin films, together with the corresponding crystallite size ($\tau$) and texture coefficient $TC_{200}$ calculated using Scherrer and (1) equations, respectively. (a) Films with varying thickness deposited at a constant $H_2$/Ar ratio of 0.08. (b) Films deposited at different $H_2$/Ar ratios with thicknesses in the range of 705–845 nm. (c) Films deposited at different $H_2$/Ar ratios with thicknesses in the range of 365–505 nm. Vertical dashed lines indicate the Bragg peak positions of cubic $\beta$-$YH_2$ (ICDD 04-002-6939).

A pronounced <100> crystallographic texture is observed for YHO films deposited at $H_2$/Ar ratio of 0.08, evidenced by the significantly higher intensity of the 200 reflection relative to the 111, 220, and 311 peaks (Fig. 2a). This texture can be interpreted within the framework of competitive grain growth in Thornton's structure zone-T [40]. Kinetic Monte Carlo simulations for *fcc* Al have demonstrated that, despite their higher surface energy, low diffusivity (100) planes can dominate sputtered films through anisotropic adatom residence times and sustained competitive overgrowth, whereas high diffusivity planes such as (111) primarily expand laterally during the early nucleation stage [41,42]. A similar growth regime appears to operate in sputtered $\beta$-$YH_x$ films, for which the (100) plane has a higher surface energy (1.60–1.83 J $m^{-2}$) than the (111) plane (0.78 J $m^{-2}$) [43]. Consistent with this interpretation, vertical fibrous grain structures with columnar grains broadening with increasing thickness – characteristic of zone-T growth – have been observed in cross-sectional SEM images of YHO films deposited under comparable conditions ($H_2$/Ar = 0.125) [25].

All samples consistently exhibit a weak but distinct 111 reflection whose intensity remains nearly constant with thickness, while the 200 reflection becomes progressively more pronounced, with $TC_{200}$ increasing from 1.3 to 2.3 (see Fig. 2a). This evolution suggests that $\beta$-$YH_x$ films initially nucleate as randomly oriented islands and subsequently develop a stronger <100> preferred orientation as the film thickness increases, consistent with the competitive grain growth mechanism discussed above.

The crystallite size shows only a weak dependence on film thickness over the investigated range (250–845 nm), with a slight tendency to decrease from approximately 12 to 10 nm with increasing thickness (Fig. 2a). This behaviour contrasts with earlier reports on YHO films, where an increase in crystallite size with increasing thickness was observed [22], suggesting differences in the film growth regime.[22]

With increasing $H_2$/Ar ratio, the pronounced <100> texture is progressively weakened ($TC_{200}$ gradually decreases) and the films evolve towards a more polycrystalline microstructure, with relative peak intensities approaching those expected for a powder diffraction pattern (see Fig. 2b,c). As discussed previously, the observed texture evolution with increasing $H_2$/Ar ratio can be explained by a growth regime transition during yttrium hydride deposition, primarily driven by changes in the hydrogen-to-yttrium arrival ratio and progressive hydridisation of the target (Fig. 1c). A slight reduction in ion-assisted growth, change in arrival particle energies, and a modest increase in total pressure act as secondary effects.

In analogy with oxygen-mediated microstructural control in polycrystalline Al films [44,45], increasing the hydrogen-to-yttrium arrival rate ratio modifies Y adatom surface mobility and grain-boundary kinetics through hydrogen adsorption, incorporation, and reaction processes, thereby influencing the overall film growth behaviour. At low hydrogen-to-yttrium ratios, zone-T competitive grain growth is preserved, resulting in a strong <100> fibre texture. At higher hydrogen fluxes, crystallite coarsening during coalescence and local epitaxial growth of individual grains are progressively disrupted, while renucleation is enhanced [46,47]. This leads to reduced crystallite sizes and a transition towards weakly textured or polycrystalline YHO films. Consistently, the crystallite size estimated from the 200 reflection using the Scherrer equation decreases systematically from 12.4 to 7.6 nm with increasing $H_2$ content (Fig. 2b,c). As microstrain broadening could not be reliably separated from size broadening, part of this decrease may also reflect changes in lattice strain or defect related broadening. As for the lattice parameter, no systematic correlation with the $H_2$/Ar ratio used during deposition is observed (see Fig. S2 in SI), except for the sample deposited at $H_2$/Ar = 1.0, which exhibits a pronounced decrease towards the $\beta$-$YH_2$ reference value.

Although film thickness was kept comparable by adjusting the deposition time, the reduced growth rate at higher $H_2$/Ar ratios increases the likelihood of interactions between the growing film and residual gas species, as well as the probability of renucleation events [44,48]. These effects further promote the formation of smaller, more randomly oriented crystallites.

### *3.3. Composition of YHO thin films*

The composition of the YHO thin films was analysed by XPS. The Y 3d spectrum contains two peaks at binding energies of 156.8 eV and 158.9 eV, corresponding to the $3d_{5/2}$ and $3d_{3/2}$ components (Fig. 3a). The position of the Y $3d_{5/2}$ peak is slightly shifted towards lower binding energy compared with $Y_2O_3$ [49], suggesting the reducing effect of hydrogen. The O 1s spectrum shown in Fig. 3b contains two peaks at binding energies of 529.2 eV, attributed to Y–O bonds, and 531.5 eV, associated with hydrocarbon contamination.

The data in Fig. 3c show a systematic decrease in oxygen content with increasing $H_2$/Ar flow ratio. Two possible interpretations may account for this behaviour. First, the reduced oxygen uptake may be associated with a lower availability of vacant lattice sites at higher hydrogen concentrations, thereby hindering subsequent oxygen incorporation and diffusion. Alternatively, the effect may originate from changes in the growth conditions that lead to differences in crystallographic texture (Fig. 2b,c), which could in turn influence film density and, consequently, oxidation kinetics. However, a pronounced increase in microstructural density with increasing $H_2$/Ar ratio appears unlikely, as film densification is typically promoted by decreasing the working pressure – i.e. under conditions of higher adatom energies – which does not seem to be the case here, as discussed above. From previous studies, it can be inferred that, in gradient YHO films, a higher oxygen content after post-oxidation is observed in <100>-textured films compared with more polycrystalline films. However, the origin of the texture gradient and the reasons why the <100> orientation promotes oxygen incorporation remain unclear [4,39].

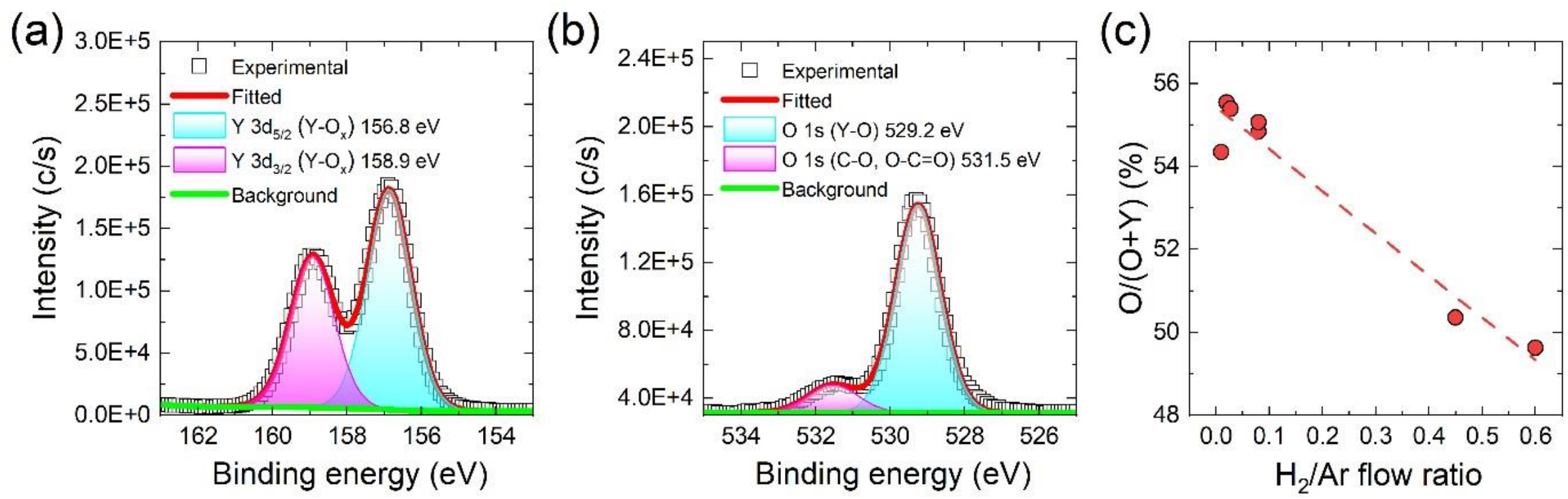

**Fig. 3.** Deconvoluted high-resolution XPS spectra of (a) Y 3d and (b) O 1s for YHO deposited at $H_2$/Ar ratio of 0.03. (c) Oxygen-to-yttrium concentration ratio, O/(O+Y), as a function of the $H_2$/Ar flow ratio during deposition.

*3.4. Optical and photochromic properties of YHO thin films*

The $T_{lum}$ and $T_{sol}$ values of YHO thin films, defined in equation (2), as a function of the $H_2$/Ar flow ratio during deposition are shown in Fig. 4a, while the corresponding transmittance spectra in the 250–2500 nm range are provided in Fig. S3 (SI). The transmittance values in Fig. 4a are not separated by film thickness, as its influence on $T_{lum}$ and $T_{sol}$ was minor compared to that of the $H_2$/Ar flow ratio. The transmittance gradually decreases from approximately 85 % to 70 % when $H_2$/Ar is increased from 0.01 to 0.60. In contrast, a sharp drop in transmittance is observed under higher hydrogen conditions ($H_2$/Ar > 0.6), reaching approximately 10 % for the film deposited at $H_2$/Ar = 1.0. For this sample, the diffraction peaks are shifted towards the $\beta$-$YH_2$ reference positions compared with the photochromic YHO films, indicating a more hydride-like character (see Fig. S4 in SI) with less oxygen concentration.

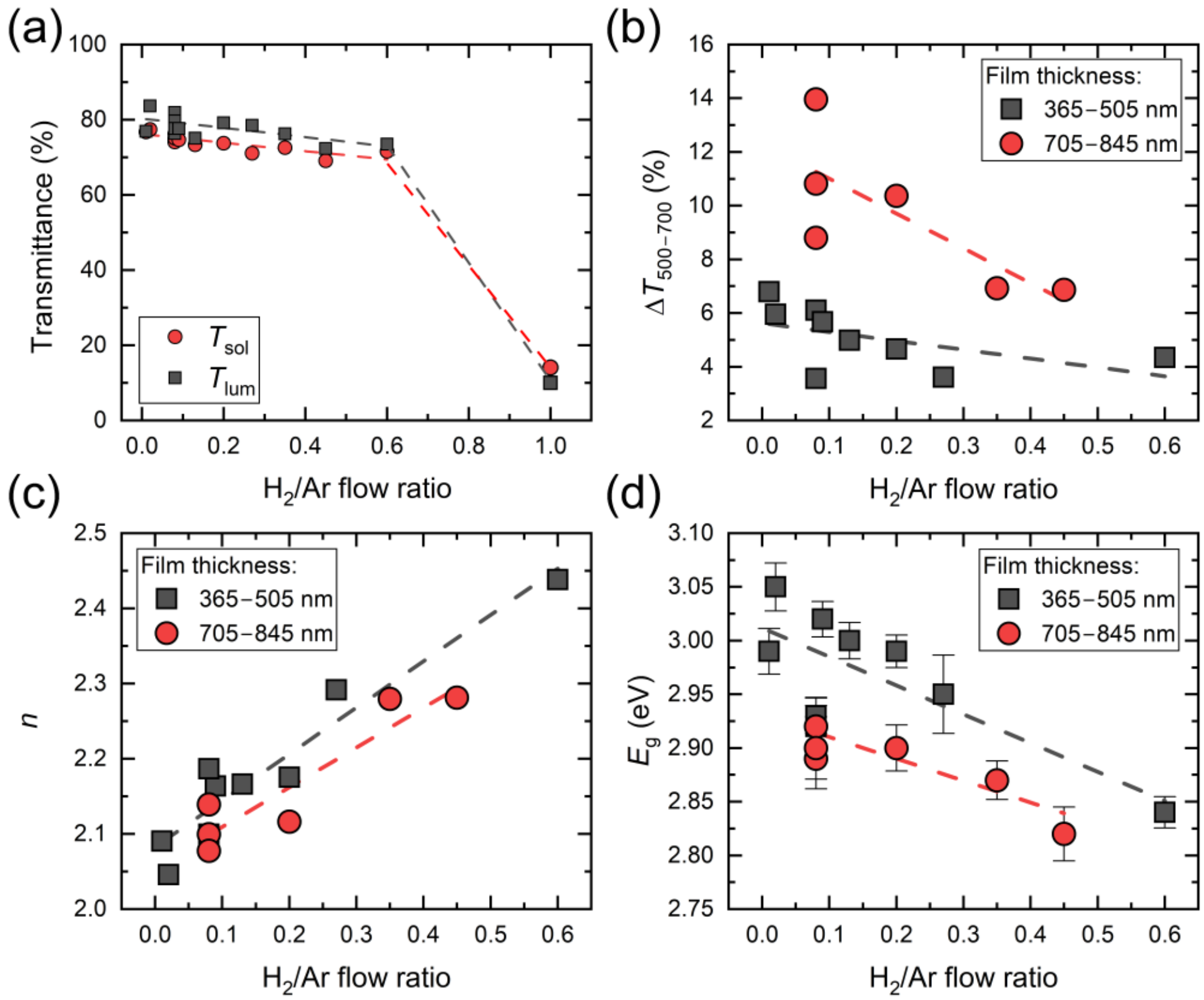

**Fig. 4.** Influence of the $H_2$/Ar flow ratio on the optical and photochromic properties of YHO thin films: (a) $T_{lum}$ and $T_{sol}$; (b) photochromic contrast in the 500–700 nm ($\Delta T_{500-700}$) range after 10 min of illumination using a solar simulator; (c) refractive index $n$ at 2.25 eV; (d) optical band gap $E_g$ derived from Tauc plots, shown separately for thinner (365–505 nm) and thicker (705–845 nm) films. Dashed lines serve as guides to the eye.

The photochromic response measurements of YHO thin films were carried out by illuminating the samples from the film surface side using a solar simulator for 10 minutes, while monitoring changes in the average optical transmittance in the 500–700 nm wavelength range (Fig. 4b). Whereas the transmittance in Fig. 4a is only marginally affected for films deposited at $H_2$/Ar flow ratios between 0.01 and 0.60, the optical contrast shows a strong dependence on this parameter, especially for thicker films, decreasing from roughly 14 % at $H_2$/Ar = 0.08 to about 7 % at $H_2$/Ar = 0.45. A similar trend, although less pronounced, most likely due to the reduced film thickness, is observed for the thinner films, with the optical contrast decreasing from about 7 % at $H_2$/Ar =

0.01 to below 5 % at $H_2/Ar \gtrsim 0.10$. The corresponding time dependent darkening and bleaching curves are presented in Fig. S5 (SI), with the transmittance normalized to its initial value to allow direct comparison of the transmittance changes between samples. The measured contrast values do not represent the complete photochromic response, because the illumination time was limited to 10 minutes and none of the films reached their fully darkened state. Consequently, films with faster switching kinetics exhibit a larger change in transmittance within the same illumination period, resulting in a higher apparent contrast.

Increased oxygen concentration in YHO thin films is generally associated with a deterioration of photochromic properties [4,39]. In the present study, however, films with higher oxygen concentrations (Fig. 3c) exhibit an enhanced photochromic response, indicating that additional factors linked to reduced hydrogen incorporation during growth may play an important role.

One possible explanation is that deposition at lower $H_2/Ar$ flow ratios is more likely to promote hydrogen-deficient $\beta$-$YH_x$ ($1.8 < x < 2.0$), resulting in incomplete occupancy of the tetrahedral anion sublattice. During subsequent oxidation, oxygen preferentially occupies the tetrahedral sites. In stoichiometric $YH_2$, this process displaces a fraction of hydrogen from tetrahedral positions into octahedral interstitial sites [1,10]. In contrast, when such sites are already available, oxygen can occupy them directly rather than substituting hydrogen, such that fewer hydrogen atoms are forced into octahedral interstitial positions. The reduced hydrogen concentration and the associated increase in unoccupied octahedral sites may modify the local electronic structure and defect landscape, potentially increasing the density of photoactive centres and thereby enhancing the optical contrast [17]. As the hydrogen content was not directly quantified in the present study, this interpretation remains a hypothesis that requires further investigation through direct measurement of the hydrogen content in films deposited at different $H_2/Ar$ flow ratios.

Both the crystallite size and the photochromic contrast vary systematically with the $H_2/Ar$ flow ratio, however, as these quantities evolve concurrently with other parameters, such as composition, a direct relationship between crystallinity and contrast cannot be established under the present deposition conditions. The role of crystallite size in determining the achievable photochromic contrast therefore warrants further investigation using approaches that enable independent control of the YHO crystallite size, for example by varying the substrate temperature during deposition.

The influence of crystallographic texture on the photochromic response is also unclear. Enhanced photochromic behaviour has recently been associated with preferential (111) orientation in YHO, whereas a pronounced (200) texture resulted in a weaker response [50]. In contrast, (200) oriented GdHO films have been reported to exhibit higher photochromic contrast than those with a (111) preferred orientation [51]. In the present study, films deposited at lower $H_2$/Ar ratios exhibit a stronger <100> texture together with generally higher photochromic contrast, with both becoming less pronounced as the $H_2$/Ar ratio increases. These contrasting findings indicate that no universal correlation between preferred orientation and photochromic performance has yet been established.

SE modelling revealed a refractive index ($n$) gradient across the film thickness, with $n$ decreasing from the substrate towards the surface, consistent with previous observations [3]. The average $n$ values at 2.25 eV are presented in Fig. 4c. For both thinner and thicker films, the refractive index shows a systematic increase with increasing $H_2$/Ar flow ratio, changing from approximately 2.05 at $H_2$/Ar = 0.01 to around 2.44 at $H_2$/Ar = 0.60. The corresponding $n$ depth profiles at 2.25 eV are shown in Fig. S6 (SI). Variations in $n$ generally arise from changes in the film's density and chemical composition [8]. However, substantial film densification is unlikely under the present deposition conditions, as discussed previously. The observed increase in $n$ is therefore more plausibly associated with compositional changes – higher H/O concentration ratio promotes a more hydride-like bonding environment with less localized electron density [38], thereby increasing the electronic polarizability of the film and consequently the refractive index [52].

The optical band gap ($E_g$), obtained from Tauc plots (see Fig. S7 in SI), as a function of the $H_2$/Ar flow ratio is shown in Fig. 4d. For both thickness ranges, $E_g$ decreases with increasing hydrogen partial pressure during deposition. In the case of thinner films, the band gap decreases from 3.05 to 2.85 eV when $H_2$/Ar ratio is increased from 0.01 to 0.60. A similar trend is observed for thicker films, with $E_g$ decreasing from approximately 2.90 eV to 2.80 eV. The higher band gap values observed at low $H_2$ partial pressures are associated with an increased oxygen content in the films, consistent with literature reports on YHO [1,4,53] and supported by our XPS measurements (see Fig. 3c).

Crucial parameter for assessing the suitability of photochromic thin films for applications such as smart windows is the time required for the material to bleach back to its initial transmittance after the illumination ceases. Previous studies on REHOs have often employed a single first-order kinetic model to describe the bleaching process [5,29,54,55], however, such an approach often does not adequately capture the full recovery behaviour observed in YHO films. Instead, the bleaching curves exhibit a clear multi-stage character, suggesting that more than one relaxation mechanism is involved. Accordingly, the experimental bleaching data were fitted using the double-exponential decay function, which in recent studies has been shown to provide a more accurate fitting of the experimental data with lower error [56,57]. From this equation, the time-dependent change in transmittance can be expressed as

$$-\ln\left(\frac{T(t)}{T_0}\right) = A_{\mathrm{b1}}\mathrm{e}^{-\frac{t}{\tau_{\mathrm{b1}}}} + A_{\mathrm{b2}}\mathrm{e}^{-\frac{t}{\tau_{\mathrm{b2}}}}, \tag{3}$$

where $T$(t) is the transmittance in time and $T_0$ represents the absolute transmittance in the fully bleached state. The bleaching process in (3) is described by two exponential components characterised by a fast time constant $\tau_{b1}$ and a slow time constant $\tau_{b2}$, with corresponding amplitudes $A_{b1}$ and $A_{b2}$. The fast component is commonly attributed to electronic transitions [13] and local structural rearrangements [17], whereas the slower component is governed by the migration of hydrogen [10] and/or oxygen-related defects [58].

To enable direct comparison between different samples using a single parameter, the two characteristic time constants were combined into the effective bleaching time constant $\tau_{b,eff}$, defined as an amplitude-weighted average,

$$\tau_{\mathrm{b,eff}} = \frac{A_{\mathrm{b1}}\tau_{\mathrm{b1}} + A_{\mathrm{b2}}\tau_{\mathrm{b2}}}{A_{\mathrm{b1}} + A_{\mathrm{b2}}}. \tag{4}$$

This parameter captures the overall bleaching response while preserving the relative contributions of the fast and slow processes and is therefore well suited for comparative analysis of bleaching behaviour across the investigated YHO thin films.

Figure 5 summarises the combined influence of the $H_2$/Ar flow ratio and film thickness on the photochromic characteristics ($\Delta T_{500-700}$ and $\tau_{b,eff}$) of YHO films for all investigated samples. When films of comparable thickness are considered, an overall trend towards higher $\Delta T_{500-700}$ and faster $\tau_{b,eff}$ is observed with decreasing $H_2$/Ar ratio, as indicated by the dashed trend lines and colour map in Fig. 5, and shown in more detail in Fig. S8(b,c) (SI). A complete overview of the film thicknesses, $H_2$/Ar ratios, $\Delta T_{500-700}$, $\tau_{b,eff}$, fitting parameters ($A_{b1}$, $\tau_{b1}$, $A_{b2}$, $\tau_{b2}$), and corresponding $R^2$ values are provided in Table S1 (SI). The individual bleaching components show that $\tau_{b1}$ remains within a relatively narrow range of 0.42–2.20 min, whereas the $\tau_{b2}$ is one to two orders of magnitude larger and varies from 7.25 to 401.85 min. Since the relative variation between $A_{b1}$ and $A_{b2}$ is considerably less, the slow component $\tau_{b2}$ contributes substantially more to the observed differences in $\tau_{b,eff}$ between the samples. The bleaching kinetics are also influenced by the achieved photochromic contrast, as demonstrated in Fig. S9 (SI). For this experiment, a separate 500 nm thick sample deposited at an $H_2$/Ar ratio of 0.01 was subjected to repeated photochromic cycling using different illumination durations to achieve varying photochromic contrasts. Each measurement was performed at a different position on the same sample to avoid the previously reported memory effect [59], and the sample was allowed to fully bleach back to its initial transmittance before the next measurement. The results show that $\tau_{b,eff}$ increases with increasing photochromic contrast, consistent with previous observations [5,55].

Film thickness represents an additional factor affecting the measured photochromic contrast. Consistent with the bulk nature of the photochromic response in YHO, thicker films generally exhibit higher optical contrast [22], as indicated by the circle size and solid vertical arrow in Fig. 5 and shown in more detail by the thickness series deposited at a fixed $H_2$/Ar ratio of 0.08 in Fig. S8(a) (SI). However, whether film thickness itself directly affects the bleaching kinetics cannot be determined from the present dataset. Given the observed dependence of $\tau_{b,eff}$ on photochromic contrast, differences in film thickness may therefore indirectly contribute to the variation in $\tau_{b,eff}$ through their effect on the achieved contrast, in addition to the influence of the $H_2$/Ar ratio.

Recent studies indicate that hydrogen diffusion in rare-earth hydrides proceeds predominantly through octahedral interstitial sites, and that higher H/Y ratios increase the occupation of these sites, thereby reducing hydrogen mobility [60]. The substantially larger variation observed for $\tau_{b2}$ compared with $\tau_{b1}$ is consistent with this interpretation, considering that the slow bleaching component has been associated with hydrogen migration [57]. In this context, a lower occupation of octahedral interstitial sites is expected to facilitate diffusion, providing a plausible explanation for the improved bleaching kinetics observed in these films [18].

Similar to the bleaching curves, the darkening curves were also approximated using a double-exponential function, yielding fast ($\tau_{d1}$) and slow ($\tau_{d2}$) components with $\tau_{d1}$ ranging from 0.05 to 0.58 min and $\tau_{d2}$ from 0.47 to 13.77 min. The resulting fitting parameters are provided in Table S2 (SI). In contrast to the bleaching measurements, which were continued until the samples approached their initial transmittance, the darkening measurements were limited to 10 min, and most samples did not reach a stable darkened state within this period. Consequently, the slow darkening component is not fully captured for these samples, resulting in a stronger dependence of the fitted $\tau_{d2}$ values on the other fitting parameters, as indicated by the high parameter dependency for most samples (Table S2, Dep. column). The $\tau_{d2}$ values should therefore be regarded as approximate, whereas the $\tau_{d1}$ occurs within the measured time window and allows a more reliable comparison between samples.

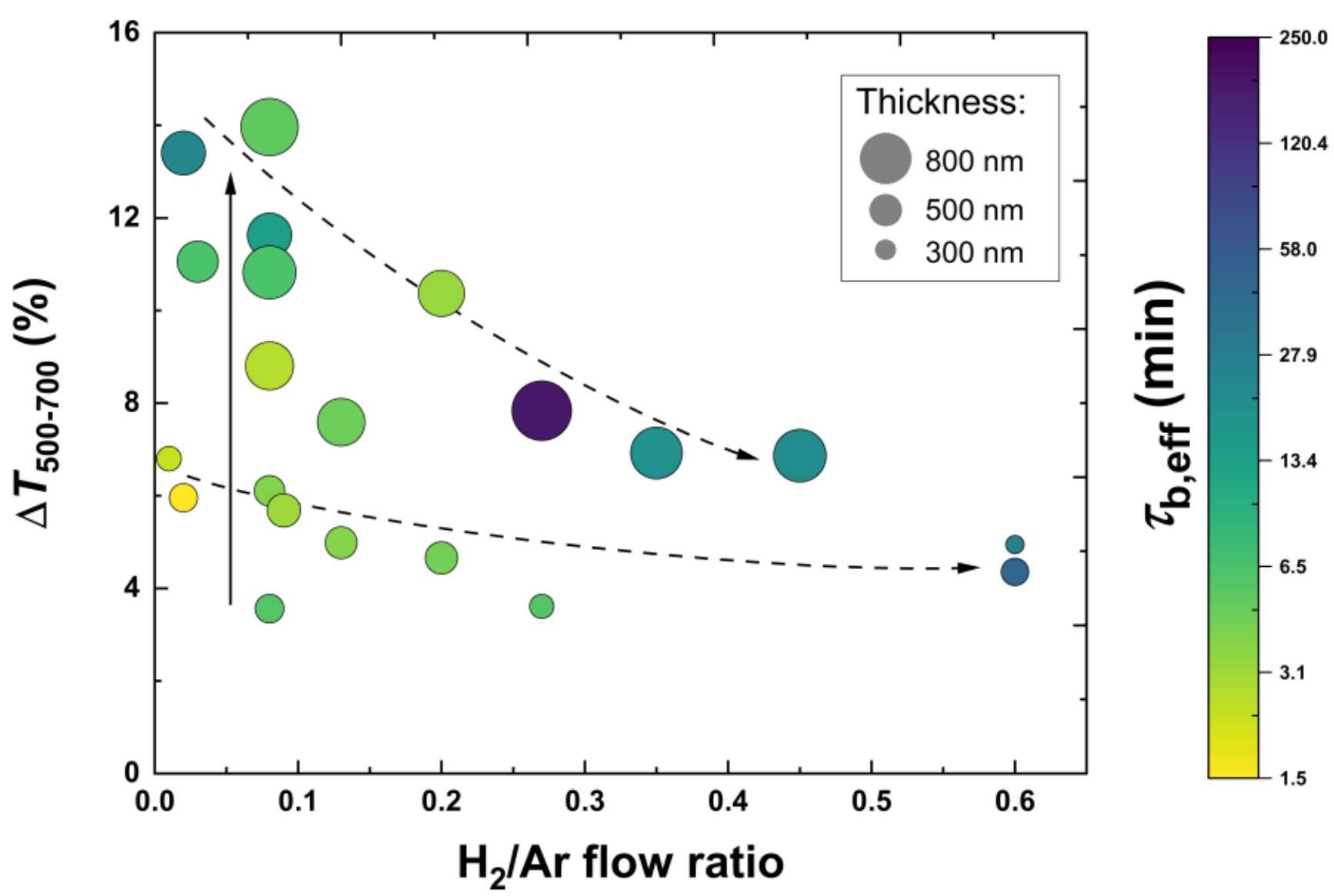

**Fig. 5.** Optical contrast $\Delta T_{500-700}$ of YHO thin films as a function of the $H_2$/Ar flow ratio used during deposition. The circle size represents the film thickness (*d*), while the colour scale indicates the effective bleaching time constant ($\tau_{b,eff}$) defined by equations (3) and (4). The solid vertical arrow highlights the increase in optical contrast with increasing film thickness at comparable $H_2$/Ar flow ratios. The dashed lines are guides to the eye, illustrating the general decrease in optical contrast with increasing $H_2$/Ar flow ratio for different thickness ranges.

Although general trends are evident, several data points deviate from the expected behaviour in Fig. 5. This variability likely stems from the strong sensitivity of YHO to subtle differences in both deposition and measurement conditions. During deposition, such differences may originate from the Y target poisoning and thus variations in thin film growth. During characterization, time delays between deposition and optical measurements can lead to differing levels of air induced oxidation.

Despite this, the results of the study show that the $H_2$ partial pressure during deposition should be kept as low as possible, while still being sufficient to form single phase $\beta$-$YH_x$ suitable to form photochromic YHO upon oxidation. Under our deposition conditions, further reduction of the $H_2$ partial pressure ($H_2/Ar < 0.01$) resulted in dark films with a mixed $\alpha$-Y/$\beta$-$YH_x$ character, consistent with the transition towards a more metallic state at low hydrogen concentrations reported in previous works [6,50]. Operating close to this threshold is favourable not only for achieving higher optical transmittance, enhanced photochromic contrast, and faster bleaching kinetics, but also from a process scaling perspective. Increasing the $H_2$/Ar ratio leads to progressive hydridisation of the Y target and an approximately 2.5-fold decrease in the deposition rate between $H_2/Ar = 0.01$ and 0.60 (see Fig. 1). Progressive target hydridisation at higher $H_2$/Ar ratios may also affect process stability during prolonged depositions. The long-term aging and cycling stability of films deposited within this operating regime, however, remains to be evaluated.

## 4. Conclusions

The influence of hydrogen partial pressure during reactive magnetron sputtering on the structure, composition, and photochromic properties of YHO thin films has been systematically investigated. The deposition process was monitored through electrical discharge parameters and in situ plasma optical emission spectroscopy, enabling direct correlation between reactive sputtering conditions and film properties.

A strong dependence of crystallographic texture and optical behaviour on the $H_2$/Ar flow ratio was observed. Lower hydrogen partial pressures promoted the formation of highly <100>-textured films with slightly larger crystallite sizes compared with the more polycrystalline structures obtained at higher $H_2$/Ar ratios. Films deposited at lower $H_2$/Ar ratios also exhibited enhanced photochromic contrast, although a direct relationship between crystallographic texture and photochromic performance cannot be established from the present results. Notably, the improved contrast was achieved despite the higher oxygen content after post-oxidation, which is typically associated with a deterioration of photochromic contrast. This suggests that the stoichiometry of the deposited $\beta$-$YH_x$ precursor, particularly its hydrogen content, plays an important role in determining the resulting optical properties of YHO. A detailed assessment of the hydrogen content before and after oxidation should therefore be addressed in future studies to establish the full range of attainable compositions via reactive magnetron sputtering.

These results demonstrate that reducing the hydrogen partial pressure during reactive sputtering – while maintaining conditions sufficient for single phase $\beta$-$YH_x$ formation – enables simultaneous improvement of key functional parameters, namely optical transparency, photochromic contrast, and bleaching speed. From a process perspective, operation at lower $H_2$/Ar ratios also provides the advantage of a higher deposition rate, which decreases by approximately a factor of 2.5 as the $H_2$/Ar ratio is increased from 0.01 to 0.60. Increasing the $H_2$/Ar ratio also leads to progressive hydridization of the Y target, as indicated by the OES and discharge characteristics, which may complicate process stability during prolonged operation. The findings highlight hydrogen partial pressure as a critical process parameter for optimising YHO thin films and provide practical guidelines for the controlled, scalable deposition of photochromic YHO films.

## Declaration of competing interest

The authors declare that they have no known competing financial interests or personal relationships that could have appeared to influence the work reported in this paper.

## Acknowledgements

Financial support was provided by Latvian Council of Science Project No. lzp-2022/1-0454. The research team acknowledges the SWEB project 101087367 funded by the HORIZON-WIDERA-2022-TALENTS-01, which provided support for competence development and training.

## Data availability

Data will be made available on request.

During the preparation of this work the author(s) used ChatGPT in order to enhance language clarity and grammatical accuracy. After using this tool/service, the author(s) reviewed and edited the content as needed and take(s) full responsibility for the content of the published article.

## References

[1] Cornelius S, Colombi G, Nafezarefi F, Schreuders H, Heller R, Munnik F, et al. Oxyhydride Nature of Rare-Earth-Based Photochromic Thin Films. Journal of Physical Chemistry Letters 2019;10:1342–8. https://doi.org/10.1021/acs.jpclett.9b00088.

[2] Mongstad T, Platzer-Björkman C, Maehlen JP, Mooij LPA, Pivak Y, Dam B, et al. A new thin film photochromic material: Oxygen-containing yttrium hydride. Solar Energy Materials and Solar Cells 2011;95:3596–9. https://doi.org/10.1016/j.solmat.2011.08.018.

[3] Zubkins M, Aulika I, Strods E, Vibornijs V, Bikse L, Sarakovskis A, et al. Optical properties of oxygen-containing yttrium hydride thin films during

and after the deposition. Vacuum 2022;203:111218. https://doi.org/10.1016/j.vacuum.2022.111218.

[4] You CC, Moldarev D, Mongstad T, Primetzhofer D, Wolff M, Marstein ES, et al. Enhanced photochromic response in oxygen-containing yttrium hydride thin films transformed by an oxidation process. Solar Energy Materials and Solar Cells 2017;166:185–9. https://doi.org/10.1016/j.solmat.2017.03.023.

[5] Zubkins M, Fedotovs A, Vibornijs V, Rudevica Z, Spunde K, Strods E, et al. Scaling up photochromic thin films with antimicrobial functionality: Roll-to-Roll deposition of YHO and YHO/Cu coatings. Vacuum 2026;244:114892. https://doi.org/10.1016/j.vacuum.2025.114892.

[6] Huiberts JN, Griessen R, Rector JH, Wijngaarden RJ, Dekker JP, de Groot DG, et al. Yttrium and lanthanum hydride films with switchable optical properties. Nature 1996;380:231–4. https://doi.org/10.1038/380231a0.

[7] Libowitz GG, Maeland AJ. Chapter 26 Hydrides, 1979, p. 299–336. https://doi.org/10.1016/S0168-1273(79)03009-9.

[8] Montero J, Martinsen FA, Lelis M, Karazhanov SZ, Hauback BC, Marstein ES. Preparation of yttrium hydride-based photochromic films by reactive magnetron sputtering. Solar Energy Materials and Solar Cells 2018;177:106–9. https://doi.org/10.1016/j.solmat.2017.02.001.

[9] Zubkins M, Gabrusenoks J, Aleksis R, Chikvaidze G, Strods E, Vibornijs V, et al. Vibrational properties of photochromic yttrium oxyhydride and oxydeuteride thin films. J Alloys Compd 2025;1015:178917. https://doi.org/10.1016/j.jallcom.2025.178917.

[10] Banerjee S, Chaykina D, Stigter R, Colombi G, Eijt SWH, Dam B, et al. Exploring Multi-Anion Chemistry in Yttrium Oxyhydrides: Solid-State NMR Studies and DFT Calculations. Journal of Physical Chemistry C 2023;127:14303–16. https://doi.org/10.1021/acs.jpcc.3c02680.

[11] Moldarev D, Primetzhofer D, You CC, Karazhanov SZ, Montero J, Martinsen F, et al. Composition of photochromic oxygen-containing yttrium hydride films. Solar Energy Materials and Solar Cells 2018;177:66–9. https://doi.org/10.1016/j.solmat.2017.05.052.

[12] Colombi G, Cornelius S, Longo A, Dam B. Structure Model for Anion-Disordered Photochromic Gadolinium Oxyhydride Thin Films. Journal of Physical Chemistry C 2020;124:13541–9. https://doi.org/10.1021/acs.jpcc.0c02410.

[13] Dam B, Nafezarefi F, Chaykina D, Colombi G, Wu Z, Eijt SWH, et al. Perspective on the photochromic and photoconductive properties of Rare-Earth Oxyhydride thin films. Solar Energy Materials and Solar Cells 2024;273:112921. https://doi.org/10.1016/j.solmat.2024.112921.

[14] Baba EM, Montero J, Strugovshchikov E, Zayim EÖ, Karazhanov EZh. Light-induced breathing in photochromic yttrium oxy-hydrides 2019. https://doi.org/10.1103/PhysRevMaterials.4.025201.

[15] Montero J, Martinsen FA, García-Tecedor M, Karazhanov SZ, Maestre D, Hauback B, et al. Photochromic mechanism in oxygen-containing yttrium hydride thin films: An optical perspective. Phys Rev B 2017;95. https://doi.org/10.1103/PhysRevB.95.201301.

[16] Chandran CV, Schreuders H, Dam B, Janssen JWG, Bart J, Kentgens APM, et al. Solid-state NMR studies of the photochromic effects of thin films of oxygen-containing yttrium hydride. Journal of Physical Chemistry C 2014;118:22935–42. https://doi.org/10.1021/jp507248c.

[17] Arslan H, Kuzmin A, Kumar Kasi V, Aulika I, Moldarev D, Primetzhofer D, et al. Anion vacancy-induced photochromism and lattice relaxation in yttrium oxyhydride. Commun Mater 2025;6:154. https://doi.org/10.1038/s43246-025-00868-2.

[18] Moldarev D, Pitthan E, Wolff M, Primetzhofer D. Effects of H vacancies on photochromic properties of oxygen-containing yttrium hydride. Nucl Instrum Methods Phys Res B 2024;555:165486. https://doi.org/10.1016/j.nimb.2024.165486.

[19] Hans M, Tran TT, Aðalsteinsson SM, Moldarev D, Moro M V., Wolff M, et al. Photochromic Mechanism and Dual-Phase Formation in Oxygen-Containing Rare-Earth Hydride Thin Films. Adv Opt Mater 2020;8. https://doi.org/10.1002/adom.202000822.

[20] Moldarev D, Wolff M, Primetzhofer D. Modification of the Photochromic Properties of Oxygen-Containing Yttrium Hydride by Irradiation with keV and MeV Ions. The Journal of Physical Chemistry C 2023;127:24676–82. https://doi.org/10.1021/acs.jpcc.3c06010.

[21] Moldarev D, Moro M V, You CC, Baba EM, Karazhanov SZ, Wolff M, et al. Yttrium oxyhydrides for photochromic applications: Correlating composition and optical response 2018. https://doi.org/10.1103/PhysRevMaterials.00.005200.

[22] Moldarev D, Wolff M, Baba EM, Moro M V., You CC, Primetzhofer D, et al. Photochromic properties of yttrium oxyhydride thin films: Surface versus bulk effect. Materialia (Oxf) 2020;11. https://doi.org/10.1016/j.mtla.2020.100706.

[23] You CC, Karazhanov SZ. Effect of temperature and illumination conditions on the photochromic performance of yttrium oxyhydride thin films. J Appl Phys 2020;128. https://doi.org/10.1063/5.0010132.

[24] Shao Z, Cao X, Zhang Q, Long S, Chang T, FangXu, et al. Fully automatic modulation for broadband sunlight based on YOxHy/VO2 hybrid structures. Solar Energy Materials and Solar Cells 2019;200:110044. https://doi.org/10.1016/j.solmat.2019.110044.

[25] Strods E, Zubkins M, Vibornijs V, Moldarev D, Sarakovskis A, Kundzins K, et al. Role of hydrogen dynamics and deposition conditions in photochromic YHO/MoO3 bilayer films. Solar Energy Materials and Solar Cells 2025;292:113789. https://doi.org/10.1016/j.solmat.2025.113789.

[26] La M, Li N, Sha R, Bao S, Jin P. Excellent photochromic properties of an oxygen-containing yttrium hydride coated with tungsten oxide (YHx:O/WO3). Scr Mater 2018;142:36–40. https://doi.org/10.1016/j.scriptamat.2017.08.020.

[27] Zhang Q, Xie L, Zhu Y, Tao Y, Li R, Xu J, et al. Photo-thermochromic properties of oxygen-containing yttrium hydride and tungsten oxide composite films. Solar Energy Materials and Solar Cells 2019;200. https://doi.org/10.1016/j.solmat.2019.109930.

[28] Chaykina D, Usman I, Colombi G, Schreuders H, Tyburska-Pueschel B, Wu Z, et al. Aliovalent Calcium Doping of Yttrium Oxyhydride Thin Films and Implications for Photochromism. The Journal of Physical Chemistry C 2022;126:14742–9. https://doi.org/10.1021/acs.jpcc.2c04456.

[29] Nafezarefi F, Cornelius S, Nijskens J, Schreuders H, Dam B. Effect of the addition of zirconium on the photochromic properties of yttrium oxy-hydride. Solar Energy Materials and Solar Cells 2019;200. https://doi.org/10.1016/j.solmat.2019.109923.

[30] Li M, Shao Z, Li Z, Zhu D, Wang J, Karazhanov SZh, et al. Co-Sputtering Crystal Lattice Selection for Rare Earth Metal-Based Multi Cation and Mixed Anion Photochromic Films. Nanomaterials 2023;13:684. https://doi.org/10.3390/nano13040684.

[31] Musil J, Baroch P, Vlček J, Nam KH, Han JG. Reactive magnetron sputtering of thin films: present status and trends. Thin Solid Films 2005;475:208–18. https://doi.org/10.1016/j.tsf.2004.07.041.

[32] Zubkins M, Sarakovskis A, Strods E, Bikse L, Polyakov B, Kuzmin A, et al. Tailoring of rhenium oxidation state in ReOx thin films during reactive HiPIMS deposition process and following annealing. Mater Chem Phys 2022;289:126399. https://doi.org/10.1016/j.matchemphys.2022.126399.

[33] Strijckmans K, Schelfhout R, Depla D. Tutorial: Hysteresis during the reactive magnetron sputtering process. J Appl Phys 2018;124. https://doi.org/10.1063/1.5042084.

[34] Harrington GF, Santiso J. Back-to-Basics tutorial: X-ray diffraction of thin films. J Electroceram 2021;47:141–63. https://doi.org/10.1007/s10832-021-00263-6.

[35] Li S-Y, Niklasson GA, Granqvist CG. Thermochromic fenestration with VO2-based materials: Three challenges and how they can be met. Thin Solid Films 2012;520:3823–8. https://doi.org/10.1016/j.tsf.2011.10.053.

[36] Mann JB. Ionization Cross Sections of the Elements Calculated from Mean-Square Radii of Atomic Orbitals. J Chem Phys 1967;46:1646–51. https://doi.org/10.1063/1.1840917.

[37] Maehlen JP, Mongstad TT, You CC, Karazhanov S. Lattice contraction in photochromic yttrium hydride. J Alloys Compd 2013;580. https://doi.org/10.1016/j.jallcom.2013.03.151.

[38] Mongstad T, Platzer-Björkman C, Karazhanov SZ, Holt A, Maehlen JP, Hauback BC. Transparent yttrium hydride thin films prepared by reactive sputtering. J. Alloys Compd., vol. 509, 2011. https://doi.org/10.1016/j.jallcom.2010.12.032.

[39] You CC, Mongstad T, Marstein ES, Karazhanov SZ. The dependence of structural, electrical and optical properties on the composition of photochromic yttrium oxyhydride thin films. Materialia (Oxf) 2019;6. https://doi.org/10.1016/j.mtla.2019.100307.

[40] Thornton JA. The microstructure of sputter-deposited coatings. Journal of Vacuum Science & Technology A: Vacuum, Surfaces, and Films 1986;4:3059–65. https://doi.org/10.1116/1.573628.

[41] Baumann FH, Chopp DL, de la Rubia TD, Gilmer GH, Greene JE, Huang H, et al. Multiscale Modeling of Thin-Film Deposition: Applications to Si Device Processing. MRS Bull 2001;26:182–9. https://doi.org/10.1557/mrs2001.40.

[42] Gilmer GH, Huang H, de la Rubia TD, Dalla Torre J, Baumann F. Lattice Monte Carlo models of thin film deposition. Thin Solid Films 2000;365:189–200. https://doi.org/10.1016/S0040-6090(99)01057-3.

[43] Tang M, Wang X, Hu X, Liu G, Liu Z, Deng H. Exploring the kinetic mechanism of dehydrogenation based on yttrium hydride surfaces: First principles calculations. Journal of Nuclear Materials 2025;614:155903. https://doi.org/10.1016/j.jnucmat.2025.155903.

[44] Barna PB, Adamik M. Fundamental structure forming phenomena of polycrystalline films and the structure zone models. Thin Solid Films 1998;317:27–33. https://doi.org/10.1016/S0040-6090(97)00503-8.

[45] Petrov I, Barna PB, Hultman L, Greene JE. Microstructural evolution during film growth. Journal of Vacuum Science & Technology A: Vacuum, Surfaces, and Films 2003;21:S117–28. https://doi.org/10.1116/1.1601610.

[46] Greene JE. Thin Film Nucleation, Growth, and Microstructural Evolution. Handbook of Deposition Technologies for Films and Coatings, Elsevier; 2010, p. 554–620. https://doi.org/10.1016/B978-0-8155-2031-3.00012-0.

[47] Pócza JF, Barna A, Barna PB, Pozsgai I, Radnóczi G. *In Situ* Electron Microscopy of Thin Film Growth. Jpn J Appl Phys 1974;13:525. https://doi.org/10.7567/JJAPS.2S1.525.

[48] Schneider JM, Larsson K, Lu J, Olsson E, Hjörvarsson B. Role of hydrogen for the elastic properties of alumina thin films. Appl Phys Lett 2002;80:1144–6. https://doi.org/10.1063/1.1448389.

[49] Cole KM, Kirk DW, Thorpe SJ. Surface Y2O3 layer formed on air exposed Y powder characterized by XPS. Surface Science Spectra 2020;27. https://doi.org/10.1116/6.0000475.

[50] Arslan H, Kuzmin A, Pudza I, Karazhanov S. Impact of hydrogen concentration on the local structure of yttrium in photochromic yttrium oxyhydride. Mater Today Chem 2026;53:103550. https://doi.org/10.1016/j.mtchem.2026.103550.

[51] Baba EM, Montero J, Moldarev D, Moro MV, Wolff M, Primetzhofer D, et al. Preferential orientation of photochromic gadolinium oxyhydride films. Molecules 2020;25. https://doi.org/10.3390/molecules25143181.

[52] Fujiwara H. Spectroscopic Ellipsometry. Wiley; 2007. https://doi.org/10.1002/9780470060193.

[53] You CC, Mongstad T, Maehlen JP, Karazhanov S. Engineering of the band gap and optical properties of thin films of yttrium hydride. Appl Phys Lett 2014;105. https://doi.org/10.1063/1.4891175.

[54] Chaykina D, Nafezarefi F, Colombi G, Cornelius S, Bannenberg LJ, Schreuders H, et al. Influence of Crystal Structure, Encapsulation, and Annealing on Photochromism in Nd Oxyhydride Thin Films. The Journal of Physical Chemistry C 2022;126:2276–84. https://doi.org/10.1021/acs.jpcc.1c10521.

[55] Colombi G, De Krom T, Chaykina D, Cornelius S, Eijt SWH, Dam B. Influence of Cation (RE = Sc, Y, Gd) and O/H Anion Ratio on the Photochromic Properties of REOxH3-2 xThin Films. ACS Photonics 2021;8:709–15. https://doi.org/10.1021/acsphotonics.0c01877.

[56] You CC, Baba EM, Nordseth Ø, Karazhanov SZh. Photochromic Properties of Yttrium Oxyhydride Powders Derived from Magnetron-Sputtered Thin Films. Physica Status Solidi (RRL) – Rapid Research Letters 2025;19. https://doi.org/10.1002/pssr.202500206.

[57] Moldarev D, Tran TT, Wolff M, Primetzhofer D. Kinetics of the photochromic effect in oxygen-containing yttrium hydrides. AIP Adv 2026;16. https://doi.org/10.1063/5.0305385.

[58] Rowberg AJE, Van de Walle CG. Defect Properties, Anion Ordering, and Photochromic Mechanism in Yttrium Oxyhydride. The Journal of Physical Chemistry C 2024;128:20387–95. https://doi.org/10.1021/acs.jpcc.4c04988.

[59] Wu Z, de Wit L, Beek M, Colombi G, Chaykina D, Schreuders H, et al. Memory effect in photochromic rare-earth oxyhydride thin films studied by *in situ* positron annihilation spectroscopy upon photodarkening-bleaching cycling. Phys Rev Mater 2024;8:045201. https://doi.org/10.1103/PhysRevMaterials.8.045201.

[60] Lin S, Jiang L, Zhang B, Luo L, Wu X, Ma Y, et al. Hydrogen diffusion in yttrium hydrides: First-principles calculations and machine learning

molecular dynamics. Int J Hydrogen Energy 2025;103:99–110. https://doi.org/10.1016/j.ijhydene.2025.01.179.